\documentclass{elsart5_mod}

 \usepackage{graphicx}

\usepackage{amssymb}

\begin{document}

\begin{frontmatter}


\title{
Kondo screening of a high-spin Nagaoka state in a triangular quantum dot 
}

\author{Akira Oguri\corauthref{cor1}}
\ead{oguri@sci.osaka-cu.ac.jp}
\corauth[cor1]{}
\author{Yunori Nisikawa}
\author{Yoshihide Tanaka}
\author{Takahide Numata}
\address{Department of Material Science, Osaka City University, 
 Sumiyoshi-ku, Osaka 558-8585, Japan}
\received{9 June 2006}
\revised{9 June 2006}
\accepted{9 June 2006}


\begin{abstract}
We study transport through a triangle triple quantum dot connected 
to two noninteracting leads using the numerical renormalization group (NRG). 
The triangle has a high-spin ground state of $S=1$ caused 
by a Nagaoka ferromagnetism, when it is isolated and 
has one extra electron introduced into a half-filling.  
The results show that the conduction electrons screen 
the local moment via two separate stages with different energy scales. 
The half of the $S=1$ is screened first by one of the channel degrees, 
and then at very low temperature the remaining half is fully screened 
to form a Kondo singlet. The transport is determined 
by two phase shifts for quasi-particles with even and odd parities, 
and then a two-terminal conductance in the series configuration is suppressed 
$g_{\rm series} \simeq 0$, while plateau of 
a four-terminal parallel conductance reaches 
a Unitary limit value $g_{\rm  parallel} \simeq 4e^2/h$ of two conducting modes.

\end{abstract}


\end{frontmatter}



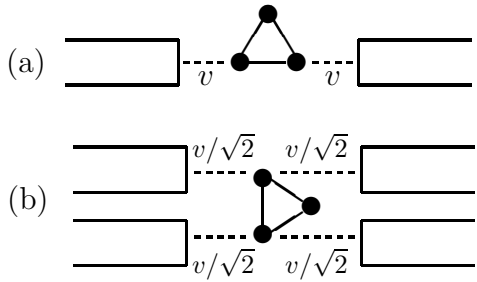
\begin{figure}[b]
\setlength{\unitlength}{0.75mm}

\rule{1cm}{0cm}
\begin{minipage}[t]{7cm}

\begin{picture}(110,20)(0,0)
\hspace{-0.8cm}
\thicklines

\put(8.5,7){\makebox(0,0)[bl]{\large (a)}}

\put(19,6){\line(1,0){20}}
\put(19,14){\line(1,0){20}}
\put(39,6){\line(0,1){8}}

\put(71,6){\line(1,0){20}}
\put(71,14){\line(1,0){20}}
\put(71,6){\line(0,1){8}}

\multiput(40.0,10)(2,0){4}{\line(1,0){1}}
\multiput(62.9,10)(2,0){4}{\line(1,0){1}}

\put(55,18.6){\circle*{3.3}} 
\put(50,10){\circle*{3.3}} 
\put(60,10){\circle*{3.3}} 

\put(51,10){\line(1,0){7}}
\put(49.5,10){\line(3,5){6}}
\put(60.5,10){\line(-3,5){6}}


\put(42.5,6){\makebox(0,0)[bl]{\large $v$}}
\put(65,6){\makebox(0,0)[bl]{\large $v$}}

\end{picture}
\end{minipage}

\setlength{\unitlength}{0.75mm}

\rule{1cm}{0cm}
\begin{minipage}[t]{7cm}

\begin{picture}(110,30)(-4,0)
\hspace{-0.8cm}
\thicklines

\put(4.75,13){\makebox(0,0)[bl]{\large (b)}}

\put(16.5,4.5){\line(1,0){20}}
\put(16.5,12.5){\line(1,0){20}}
\put(36.5,4.5){\line(0,1){8}}

\put(16.5,17.5){\line(1,0){20}}
\put(16.5,25.5){\line(1,0){20}}
\put(36.5,17.5){\line(0,1){8}}

\put(67.5,4.5){\line(1,0){20}}
\put(67.5,12.5){\line(1,0){20}}
\put(67.5,4.5){\line(0,1){8}}

\put(67.5,17.5){\line(1,0){20}}
\put(67.5,25.5){\line(1,0){20}}
\put(67.5,17.5){\line(0,1){8}}

\multiput(38.0,21)(2,0){5}{\line(1,0){1}}
\multiput(38.0,9.5)(2,0){5}{\line(1,0){1}}
\multiput(53.2,9.5)(2,0){7}{\line(1,0){1}}
\multiput(53.2,21)(2,0){7}{\line(1,0){1}}

\put(58.6,15){\circle*{3.3}} 
\put(50,10){\circle*{3.3}} 
\put(50,20){\circle*{3.3}} 

\put(50,10){\line(0,1){10}}
\put(51,10){\line(3,2){6}}
\put(51,20){\line(3,-2){6}}


\put(37.5,22.5){\makebox(0,0)[bl]{$v/\sqrt{2}$}}
\put(54,22.5){\makebox(0,0)[bl]{$v/\sqrt{2}$}}

\put(37.5,2){\makebox(0,0)[bl]{$v/\sqrt{2}$}}
\put(54,2){\makebox(0,0)[bl]{$v/\sqrt{2}$}}

\end{picture}
\end{minipage}
\caption{Schematic picture of (a) series and (b) parallel connections}
\label{fig:system}
\end{figure}

The Kondo effect in quantum dots 
is an active field of current research,
and recently triple quantum dots with 
triangle have been examined intensively \cite{Vidan-Stopa,TKA}.  

One interesting property expected to be seen in a quantum-dot array  
with closed paths is that some degenerate states 
could be lifted by circular orbital motions of 
electrons to form a high-spin ground state 
due to the Nagaoka mechanism.
In this report for clarifying, $i$) how the Nagaoka ferromagnetism 
that could manifest in the isolated triangle 
for a particular charge filling is screened 
by the conduction electrons, and 
$ii$) how it affects the low-temperature transport bellow 
the Kondo energy scale $T_K$, 
 we present the results using 
the NRG approach \cite{NO,ONH,OH}, 
which is applicable to low-temperatures $T \lesssim T_K$.

We start with a three-site Hubbard model 
 connected to two non-interacting leads 
on the left($L$) and right($R$), as 
shown in Fig.\ \ref{fig:system} (a):  
$\quad \mathcal{H} =  \mathcal{H}_{D} +  \mathcal{H}_{\rm mix}  +  
\mathcal{H}_{\rm lead} $, 
\begin{eqnarray}
\mathcal{H}_{D}  \,=\,   
 -t 
 \sum_{<ij>}^{N_D}\sum_{\sigma}  
 \left(\,
 d^{\dagger}_{i\sigma}d^{\phantom{\dagger}}_{j\sigma}  
+ d^{\dagger}_{j\sigma}d^{\phantom{\dagger}}_{i\sigma}  
\right) 
\rule{1.7cm}{0cm} 
\nonumber
 \\
+\,
\epsilon_d 
\sum_{i=1}^{N_D}\sum_{\sigma} \, 
 d^{\dagger}_{i\sigma}d^{\phantom{\dagger}}_{i\sigma}  
 +  U\sum_{i=1}^{N_D} 
   d^{\dagger}_{i \uparrow}
   d^{\phantom{\dagger}}_{i \uparrow}
   d^{\dagger}_{i \downarrow}
   d^{\phantom{\dagger}}_{i \downarrow} \;,
 \rule{0.2cm}{0cm} 
\label{eq:HC^U}
\\
\mathcal{H}_{\rm mix} \,=\, 
v  \sum_{\sigma}   
 \left(  \,
  d^{\dagger}_{1,\sigma} \psi^{\phantom{\dagger}}_{L \sigma}
   +   
d^{\dagger}_{N_D, \sigma} \psi^{\phantom{\dagger}}_{R\sigma}
+\, \mathrm{H.c.}
\, \right)   ,
\label{eq:Hmix}
 \rule{0.3cm}{0cm} 
\\
\mathcal{H}_{\rm lead}  \,=\, 
\sum_{\nu=L,R} 
 \sum_{k\sigma} 
  \epsilon_{k \nu}^{\phantom{0}}\,
         c^{\dagger}_{k \nu \sigma} 
         c^{\phantom{\dagger}}_{k \nu \sigma}
\,,
\rule{2.6cm}{0cm} 
\label{eq:H_lead}
\end{eqnarray}
where $t$ is the hopping matrix element between the dots, 
 $\epsilon_d$ the onsite energy, $U$ the Coulomb interaction,  
and $N_D =3$. 
A linear combination of the conduction electrons 
 $\psi_{\nu \sigma}^{\phantom{\dagger}} 
\equiv \sum_k c_{k \nu \sigma}^{\phantom{\dagger}}/\sqrt{N}$ 
hybridizes with the electrons in the dots via $v$, 
or $\Gamma \equiv \pi v^2 \rho$, 
where $\rho$ is the density of states for each lead.
The low-energy states of the whole system including the leads 
show a local Fermi-liquid behavior, 
which is characterized by two phase 
shifts $\delta_{\rm even}$ and $\delta_{\rm odd}$ 
for the quasi-particles with the even and odd parities. 
Then the dc conductance  $g_{\rm series}$ and 
total number of electrons in the dots  
$N_{\rm el}$ 
can be expressed at $T=0$ in the form \cite{NO,ONH},
\begin{eqnarray}
g_{\rm series} \,=\, \frac{2e^{2}}{h}\,
\sin^{2} \Bigl(\,\delta_{\rm even}
-\delta_{\rm odd}\,\Bigr)\;,
\rule{2cm}{0cm}
\label{eq:gs}
\\
N_{\rm el}
\equiv \,\sum_{i=1}^{N_D}\sum_{\sigma} \,
\langle d_{i\sigma}^{\dagger}d_{i\sigma}^{\phantom{\dagger}}\rangle
\, = \, \frac{2}{\pi}\Bigl(\,\delta_{\rm even}+\delta_{\rm odd}\,\Bigr)\;.
\label{eq:N_el}
\end{eqnarray}
Furthermore, because of some symmetrical reasons, 
the conductance for the current flowing in the horizontal direction 
in a four-terminal geometry, as shown in Fig.\ \ref{fig:system} (b), 
can be related to these two phase shifts defined with respect to 
the series connection \cite{NO,ONH},
\begin{equation}
g_{\rm  parallel} \,= \,\frac{2e^{2}}{h} \,
 \Bigl(\,\sin^{2}\delta_{\rm even} \, + \,
 \sin^{2}\delta_{\rm odd}\,\Bigr)\;.
      \label{eq:gp}
\end{equation}
We have deduced $\delta_{\rm even}$ and $\delta_{\rm odd}$ from the 
fixed-point eigenvalues of the discretized Hamiltonian $H_N$ of NRG.

  \begin{figure}[t]
\begin{center}
 \leavevmode
\includegraphics[width=0.88\linewidth]{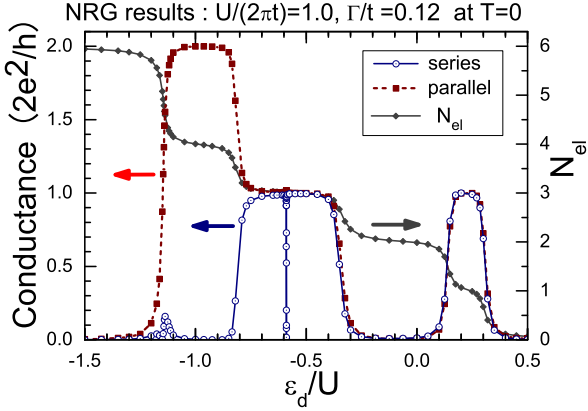}
 \caption{
Series ($\odot$) $g_{\rm  series}$ and parallel 
 ({\protect\small $\blacksquare$}) $g_{\rm  parallel}$ conductances  
are plotted together with local charge ($\blacklozenge$) $N_{\rm el}$ 
as functions of $\epsilon_d$ for $U/(2\pi t)=1.0$ and $\Gamma/t = 0.12$.
In NRG, $t/D =0.1$ and $\Lambda=6.0$.
 }
 \label{fig:cond_u1}
 \end{center}
  \end{figure}

In Fig.\  \ref{fig:cond_u1}, 
the results of $\,g_{\rm  series}$, $\,g_{\rm  parallel}$  
and $N_{\rm el}$ are 
plotted as functions of $\epsilon_d$ that corresponds 
to the gate voltage. 
The number of electrons inside the triangle increases 
with decreasing $\epsilon_d$ showing 
a staircase behavior at $N_{\rm el} \simeq1,2,3,4,$ and $6$.
Due to a degeneracy arising in the ground state 
at $\epsilon_d/U \simeq -1.15$, 
the average charge changes directly form $\,4$ to $\,6$ without 
taking a step corresponding to $N_{\rm el} \simeq 5$. 
We see that the conductances for small filling  $\epsilon_d/U \gtrsim -0.8$ 
show the typical Kondo behavior, which has also been seen 
in a linear chain of quantum dots \cite{NO,ONH}. 
Namely, for odd occupancies $N_{\rm el} \simeq 1$ and $3$, 
the plateaus of $\,g_{\rm  series}$ and $\,g_{\rm  parallel}$ reach 
 the Unitarly-limit value $2e^2/h$ of a single conduction mode.
We also found a very narrow dip of $\,g_{\rm  series}$ at 
$\epsilon_d/U \simeq -0.6$.
For an even occupancy at $N_{\rm el} \simeq 2$, the conductances are suppressed.The irregular behavior at $N_{\rm el} \simeq 4$ is 
partly relating to the fact that, for $U=0$, the circular motion 
along the triangle forms two degenerate orbitals 
at $\varepsilon_b \equiv t$ and a single orbital  
at $\varepsilon_a \equiv -2t$. As two of the four electrons  
occupy the $\varepsilon_b$ orbitals, the Coulomb repulsion 
lifts the degeneracy and makes a $S=1$ state 
to be a many-body ground state for $\Gamma=0$. 
This corresponds to the Nagaoka state, 
and also links to a flat-band ferromagnetism.  
Then, the conduction electrons from the leads screen 
the moment. At the filling $N_{\rm el} \simeq 4$,
the phase shifts take the values $\delta_{\rm even} \simeq 3\pi/2$ 
and $\delta_{\rm odd}\simeq \pi/2$. 
It suppresses the series conductance $g_{\rm  series} \simeq 0$, 
while for the parallel conductance both of 
the two conducting modes contribute to 
the plateau of $\,g_{\rm  parallel} \simeq 4e^2/h\,$ at $T=0$. 
The difference between the peak values of 
the two conductances itself is caused by an interference effect. 

\begin{figure}[tb]
\begin{center}
\leavevmode
\includegraphics[width=0.8\linewidth]{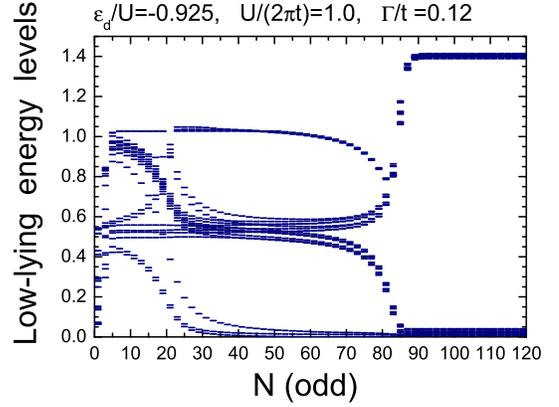}
\caption{
 Low energy eigenvalues of a discretized Hamiltonian $H_N/D$  
as functions of odd NRG step $N$ for $\epsilon_d/U= -0.925$,
where  $N_{\rm el} \simeq 4$.
Other parameter are taken to be the same 
as those used for Fig.\ \ref{fig:cond_u1}. 
}
\label{fig:flow}
\end{center}
\end{figure}

In order to give some insights into the screening mechanism of
the $S=1$ moment, the low-lying energy levels of 
the NRG version of Hamiltonian $H_N$ are plotted against odd $N$ 
in Fig.\ \ref{fig:flow}. 
The trajectory of the levels shows how 
the system evolves from the high-energy regime 
to the low-energy Fermi-liquid regime as $N$ increases,
or equivalently with decreasing energy 
 $\omega_N \sim D \Lambda^{-(N-1)/2}$,   
where $D$ is the half-width of the conduction bands 
and $\Lambda$ is the parameter for the logarithmic discretization. 
Our results indicate clearly that two crossovers occur separately 
at $N \simeq 20$ and $N \simeq 80$. 
This feature and additional information about 
the total spin and electron number of the eigenstates
show that at high temperatures $T \gtrsim \omega_{N=20}$   
the $S=1$ moment is free from the screening.
Then at the intermediate temperatures, 
 $\omega_{N=80} \lesssim T \lesssim \omega_{N=20}$, 
the half of the moment is screened by the conduction electrons from 
one of the channel degrees, 
and thus in this region the local moment is still 
in an under-screened situation. 
The full screening is completed 
finally at very low temperatures $T \lesssim \omega_{N=80}$.
 
The competition between the Kondo effect and Nagaoka ferromagnetism 
could happen in a wide of class of quantum dots  
that have closed paths for the orbital motion.


This work was supported 
by the Grant-in-Aid for Scientific Research from JSPS.
Numerical computation was partly carried out 
in Yukawa Institute Computer Facility.

\end{document}